\documentclass[lettersize,journal]{IEEEtran}
\usepackage{amsmath,amsfonts}
\usepackage{algorithmic}
\usepackage{algorithm}
\usepackage{array}
\usepackage[caption=false,font=normalsize,labelfont=sf,textfont=sf]{subfig}
\usepackage{textcomp}
\usepackage{stfloats}
\usepackage{url}
\usepackage{verbatim}
\usepackage{multirow}
\usepackage{graphicx}
\usepackage{cite}
\usepackage{makecell}
\usepackage{array} 
\usepackage{threeparttable}
\usepackage{enumitem}
\usepackage{csquotes}
\usepackage[colorlinks,
linkcolor=red,
anchorcolor=blue,
citecolor=green
]{hyperref}

\begin{document}

\title{GSsplat: Generalizable Semantic Gaussian Splatting for Novel-view Synthesis in 3D Scenes} 

\author{Feng Xiao, Hongbin Xu, Wanlin Liang, Wenxiong Kang 
}

\maketitle

\begin{abstract}
The semantic synthesis of unseen scenes from multiple viewpoints is crucial for research in 3D scene understanding. Current methods are capable of rendering novel-view images and semantic maps by reconstructing generalizable Neural Radiance Fields. However, they often suffer from limitations in speed and segmentation performance. We propose a generalizable semantic Gaussian Splatting method (GSsplat) for efficient novel-view synthesis. Our model predicts the positions and attributes of scene-adaptive Gaussian distributions from once input, replacing the densification and pruning processes of traditional scene-specific Gaussian Splatting. In the multi-task framework, a hybrid network is designed to extract color and semantic information and predict Gaussian parameters. To augment the spatial perception of Gaussians for high-quality rendering, we put forward a novel offset learning module through group-based supervision and a point-level interaction module with spatial unit aggregation. When evaluated with varying numbers of multi-view inputs, GSsplat achieves state-of-the-art performance for semantic synthesis at the fastest speed. 
\footnote{The code will be released at \href{https://github.com/onmyoji-xiao/GSsplat}{{https://github.com/onmyoji-xiao/GSsplat}}.} 
\end{abstract}


\section{Introduction}
Dense semantic mapping of scene geometry is a fundamental aspect of 3D scene understanding. It enhances the ability to interpret complex environments, crucial for applications such as visual navigation and robotic interaction\cite{lei2023recent}. Semantic view synthesis has received massive attention in this field. Due to the limitations of the discrete 3D representation by point clouds, meshes, and voxel \cite{he2021deep} on view synthesis, recent works focus more on extending semantic reasoning from radiance field approaches \cite{nguyen2024semantically}.

Neural Radiance Fields (NeRF) \cite{mildenhall2021nerf} is an implicit volumetric representation that captures complex lighting and geometric details, enabling high-quality 3D reconstructions with fewer images. Semantic NeRF generally synthesizes semantic views by integrating an additional segmentation renderer. Given source views and camera parameters, the color and volume density of sampled points on rays from multi-layered perceptions (MLPs) are accumulated to create pixel-wise semantic labels \cite{zhi2021place}. 3D Gaussian Splatting (3DGS) \cite{kerbl20233d} is another radiance field approach as explicit scene representation, constructing millions of learnable 3D Gaussians to achieve real-time rendering \cite{chen2024survey}. Similarly, semantic features are embedded in color functions of 3D Gaussian points to get novel-view semantic class labels \cite{guo2024semantic}. Nevertheless, these approaches all entail the establishment of radiance fields specific to the scene, involving exorbitant training expenses, and cannot be generalized to unseen scenes. 
\begin{figure}[!t]
        \centering{\includegraphics[width=0.9\linewidth]{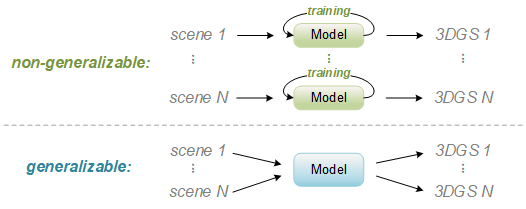}}
	\caption{The top is the non-generalizable 3DGS reconstruction process and the bottom is the generalizable inference.}
   \vspace{-0.4cm}
    \label{fig_start}
\end{figure}

Existing methods have implemented generalizable NeRF for novel-view semantic synthesis in indoor scenes \cite{liu2023semantic, chen2023gnesf, chou2024gsnerf}, but still exhibit notable deficiencies. Firstly, the majority of semantic radiance fields are overly dependent on the segmentation effect of 2D images during the construction process, ignoring the semantic analysis related to spatial information. Alternatively, NeRF-based methods possess a considerable supremacy in the image rendering quality, but consume a lot of time due to the calculation of dense sampling points. The latest 3DGS-based method with semantic synthesis uses a generalizable Gaussian feature extraction network \cite{jurca2024rt} but still requires the prior construction of scene-specific Gaussian radiance fields. As shown in Figure \ref{fig_start}, the non-generalizable method requires iterative training in each scene to reconstruct the Gaussian radiation field, while the generalizable method only requires one forward inference of the model to predict the Gaussian distributions of different scenes. MVsplat\cite{chen2025mvsplat} constructs Gaussian points from the depth information of multiple views without requiring training for densification and pruning in a specific scene. However, it only realizes generalizable 3DGS in color rendering, the network that relies heavily on cost volume supervision cannot be directly migrated to semantic understanding. 

We aim to tackle the aforementioned issues by extending multi-task view synthesis into generalizable Gaussian Splatting, while facing several challenges. When the Gaussian distributions are insufficient to fully represent a scene, 3DGS achieves high-quality rendering via an adaptive density control mechanism\cite{liu2024georgs}. Obviously, it cannot be generalized to other scenes. Previous works use depth priors to guide the geometric structure in the radiance field\cite{guo2024depth, zhu2025fsgs}, is it possible to generate Gaussian point clouds directly by depth? In fact, due to the effects of light and noise, the corresponding points in different views are not always the same in terms of brightness and texture\cite{huang2024multi}. 3DGS generates geometrically independent points to fit the scene to achieve high-quality rendering during the iterative process of densification. Thus, predicting Gaussians directly from multi-view inputs in one inference remains challenging. Furthermore, semantics and appearance exhibit significant differences in the feature space. Simultaneously predicting two Gaussian parameters in an end-to-end network is likely to influence each other.  

In this paper, we propose GSsplat, a generalizable semantic Gaussian Splatting method for efficient novel-view synthesis in 3D scenes. It reconstructs Gaussian radiance fields from multi-view inputs without scene-specific training, and is capable of rendering both color images and semantic maps. We devise a hybrid network for extracting pixel-level color and semantic features, and subsequently predict Gaussian distributions on the point cloud initialized by depth priors. We propose two innovative modules to optimize Gaussian attributes and their positions, enhancing scene-fitting capabilities instead of relying on densification and pruning iterations in non-generalizable methods. A point-level interaction module with feature aggregation is introduced, enabling the Gaussian distribution prediction process to effectively learn the cross-view information for parameter reasoning. Distinct from other generalizable 3DGS-based methods, we conduct group-supervised learning on the original Gaussian points considering the offset conditions. The points with offsets are utilized to adapt to the scene to attain high-quality rendering, while the remaining points keep the basic geometry. 

The main contributions of our work can be summarized as follows:
\begin{itemize}[leftmargin=*]
\item{We extend the novel-view synthesis with semantic understanding to 3DGS and propose a generalizable Gaussian Splatting method, which enables Gaussian radiance field reconstruction in unseen scenes with color and semantic rendering.}
\item{We design a universal hybrid network to enhance the understanding and reconstruction of scene semantics in the multi-task Gaussian framework, achieving much better results than traditional single-encoder structures.}
\item{Distinct from the existing Gaussian prediction approaches, we present a novel offset learning module that employs geometric constraints to group-supervise the Gaussian positions. And the point-level interaction module significantly boosts the predictive capability of parameters.}
\item{Our GSsplat attains state-of-the-art performance in novel-view semantic synthesis and holds the fastest feed-forward speed among multi-task rendering frameworks.}
\end{itemize}

\section{Related Work}
\subsection{Semantic Radiance Field}
NeRF-based methods with implicit representation are essential for 3D scene reconstruction and understanding. Semantic-NeRF \cite{zhi2021place} is a pioneering effort that formalizes semantic segmentation as an inherently view-invariant function to append a segmentation renderer. NeSF \cite{vora2021nesf} builds on the pre-trained NeRF and employs a 3D UNet for extended semantic training across various scenes. Most of the subsequent work relies on these two frameworks for semantic rendering. Panoptic NeRF \cite{fu2022panoptic} points out that inaccurate geometric reconstruction leads to wrong semantic maps, and optimizes it by 3D-to-2D label transfer. OR-NeRF \cite{yin2023or} introduces a point projection strategy to find correspondences from 2D pixels to 3D point clouds, ensuring 3D consistency with less processing burden. Otherwise, some work \cite{kerr2023lerf, zhang2024open, wang2024gov, 10630553} combine visual-language models with NeRF to achieve open-vocabulary semantic segmentation.

Unlike implicit modeling of NeRF, 3DGS is an emerging technology with explicit 3D representation that enables real-time rendering of complex scenes. Applying 3DGS to 3D scene understanding has become an up-and-coming trend \cite{wu2024recent}. Gaussian Grouping \cite{ye2023gaussian} represents the whole 3D scene with a set of grouped 3D Gaussians which lifts knowledge of the pre-trained 2D segmentation model. Another more efficient model \cite{lan20232d} employs 2D semantics to supervise Gaussian rendering, while nearest neighbor clustering and statistical filtering refine the segmentation results. 3DGS are also popularly used in open-vocabulary 3D scene segmentation by projecting semantic features from pre-trained 2D encoders \cite{guo2024semantic, shi2024language, zhou2024feature}. LangSplat \cite{qin2024langsplat} leverages SAM \cite{kirillov2023segment} to learn hierarchical semantics and map multi-view embeddings to a low-dimensional latent space, while OpenGaussian \cite{wu2024opengaussian} highlights the limitations of 2D feature maps in handling occlusion and proposes enhanced point-level understanding.

\subsection{Generalizable novel-view synthesis}
The recent novel-view synthesis methods, NeRF and 3DGS, require specific training for each scene to reconstruct it accurately. Generalizing them to new scenes is a significant challenge in practical applications \cite{bao20243d}. In NeRF research, scene-independent inference is typically achieved by incorporating structures that aggregate reference-view features or parameterizing a target network to modify implicit representations for different scenes \cite{yu2021pixelnerf, ye2023featurenerf, bao2023insertnerf}. However, 3DGS relies on establishing the initial set of sparse points and the non-differentiable operations of point densification and pruning, which is not conducive to generalization. With the development of multi-view stereo \cite{li2023hybrid, xu2024robustmvs}, the current methods initialize the spatial coordinates by view consistency and then predict the parameters for each Gaussian. PixelSplat \cite{charatan2024pixelsplat} designs a multi-view epipolar transformer for 3D reconstruction and regards training Gaussian primitives as differentiable dense probability learning. MVsplat \cite{chen2025mvsplat} suggests creating a cost volume representation to predict the centers of Gaussians. LatentSplat \cite{wewer2024latentsplat} introduces variational 3D Gaussians, which can explicitly model uncertainty by holding semantic feature distributions in object-centric scenes. 

Although these methods successfully implement novel-view synthesis with generalization, scene understanding typically requires semantic reasoning when rendering novel views. Existing methods have accomplished the synthesis of novel-view images and semantic segmentation in a generalizable NeRF \cite{liu2023semantic, chen2023gnesf, chou2024gsnerf}, but the speed is unsatisfactory. Some work like RT-GS2 \cite{jurca2024rt} performs a second optimization with 3DGS to enhance its generalization ability but still relies on pre-trained single-scene models. Therefore, we focus on generalizable 3DGS without prior modeling to achieve multi-task view synthesis of 3D scenes.

\begin{figure*}[t]
        \centering{\includegraphics[width=1.0\linewidth]{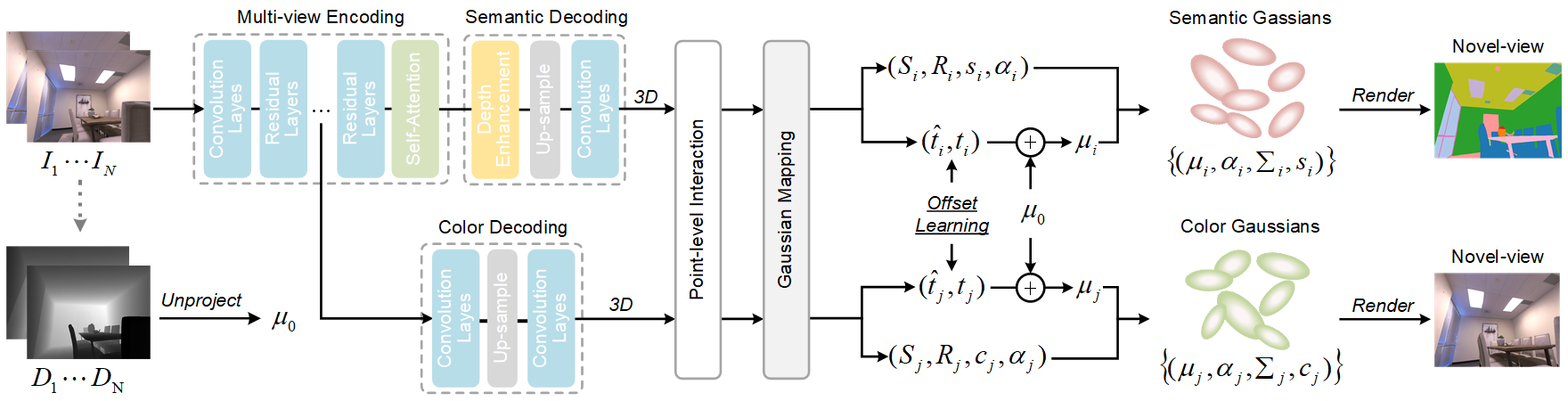}}
	\caption{\textbf{Overview of GSsplat.} Given $N$ source views and camera poses, our model directly predicts the semantic Gaussian and color Gaussian parameters from RGB and depth information for 3D scene reconstruction. Firstly, the hybrid network uses a multi-view encoding module to extract 2D semantic and color features. Next, the features are decoded to the original image resolution and unprojected per pixel to the 3D space. After point-level interaction and Gaussian mapping, the semantic and color Gaussian radiance fields are constructed by the predicted parameters, and novel views are rendered through the Gaussian splatting operation.}
   \vspace{-0.4cm}
    \label{fig_overall}
\end{figure*}

\section{Methods}

\subsection{Fundamentals}
3DGS is an explicit radiance field technique based on Gaussian distribution for 3D scene representation and rendering. It typically models each scene, requiring dense images of the scene with different poses to optimize a set of learnable Gaussian functions\cite{dalal2024gaussian}. $G(x)=exp{-\frac{1}{2}(x-\mu)^T\Sigma^{-1}(x-\mu)}$ denotes the Gaussian distribution at the spatial coordinates $x$ with covariance matrix $\Sigma$ and mean $\mu$. In the radiance field, $\Sigma$ is reparameterized as a combination of the rotation matrix $R$ and the scaling matrix $S$: $\Sigma=RSS^TR^T$. 3DGS projects Gaussian primitives onto a 2D image plane using a rasterization renderer. The Gaussian opacity $\alpha$, the color function $f(\theta, \phi)$ and $G(x)$ are computed to effectuate the rendering under the viewing direction $(\theta, \phi)$. The rendering process $Render(x,\theta, \phi)=\sum_i \alpha_if_i(\theta, \phi)G_i(x)$ is essentially a blending of all Gaussian contributions from a specific viewpoint. In the most recent research, $f(\theta, \phi)$ is not merely delineated by the spherical harmonics coefficients for RGB values, but can also represent feature vectors with any other connotations and dimensions. In this paper, we use it to establish two kinds of 3D Gaussians for novel-view synthesis, corresponding to color rendering and semantic segmentation. 

\subsection{Multi-task Framework}
As shown in Figure \ref{fig_overall}, we construct explicit 3D scene representations from multi-view inputs through a 3DGS-based multi-task framework. The generalization involves directly predicting the 3D Gaussian parameters from the learnable pixel-level features and the mapping relations between 2D pixels and 3D points. Our goal is to extract both color and semantic information using an end-to-end network and output two 3D Gaussian radiance fields for rendering and synthesis. Unlike existing radiance field approaches that use a single structure to learn multi-view semantics and colors together, we design a hybrid network to account for the inherent differences between color and semantics. 

\subsubsection{Multi-view Encoding}
In the encoding process, input images $I \in\mathbb R^{K\times H\times W \times 3}$ of $K$ views are fed into the network consisting of residual layers and self-attention layers. In contrast to high-level semantic understanding, the perception of color relationships is more straightforward and necessary to prevent the blurring from overly deep convolution. Therefore, we allow the color and semantic encoding to share the shallow layers, while deeper layers learn more complex semantic features, which boosts efficiency and enhances the disparities between them in the feature space. Then we obtain color features $F_r \in\mathbb R^{\frac{H}{2}\times \frac{W}{2} \times C_e}$ and semantic features $F_s \in\mathbb R^{\frac{H}{4}\times \frac{W}{4} \times C_e}$.

\subsubsection{Semantic and Color Decoding}
The output $F_r$ and $F_s$ are decoded via distinct branches and up-sampled to the original image resolution. To enhance the semantic comprehension from geometric relationships, we incorporate the scaled depth maps $D\in\mathbb R^{K \times \frac{H}{4}\times \frac{W}{4}}$ into the encoded semantic features, mitigating the perceptual limitations of 2D vision at the object level. The decoding module eventually yields feature vectors $F'_r \in\mathbb R^{H\times W \times C_d}$ and $F'_s \in\mathbb R^{H\times W \times C_d}$ for pixel-level color and semantic expressions. Subsequently, both are mapped to Gaussian parameters to reconstruct 3D radiance fields. 

\subsection{Generalizable Gaussians}
The generalizability of 3DGS is reflected in its ability to predict Gaussian parameters by extracting 3D scene information from multiple views without densification and pruning iterations in a specific scene. Similar to the non-semantic generalizable 3DGS-based methods\cite{chen2025mvsplat, szymanowicz2024flash3d}, we use deep priors to initialize inaccurate Gaussian point clouds. The pixel-level features are mapped to 3D space as the expression of Gaussian distributions. Nevertheless, multi-view features are extracted separately for each image, only considering Gaussian-related attributes under single-view information, lacking interaction of different views. The previous approach in \cite{chen2025mvsplat} employs cross-view attention to attain information interaction, yet disregards the geometric relationships at the 3D level. 

Additionally, to acquire high-quality renderings at arbitrary angles, the positions of the Gaussian distribution fitted therein are typically not completely matched with the stereoscopic geometry of the visible views. In the generalizable Gaussian reconstruction process, it is necessary to further learn scene-adaptive Gaussian distributions to replace the traditional densification and pruning. The Gaussian distribution initialized by deep priors obviously cannot represent the final positions. The existing generalizable approaches improve the rendering quality by learning shift-Gaussian or multi-Gaussian \cite{szymanowicz2024flash3d}, but only carrying out reconstruction in the depth or pixel dimensions instead of directly correlating with 3D positions. Therefore, we propose a point-level interaction module by spatial feature aggregation and a novel Gaussian offset prediction module with geometric constraints. 

\begin{figure}[b]
        \centering{\includegraphics[width=1.0\linewidth]{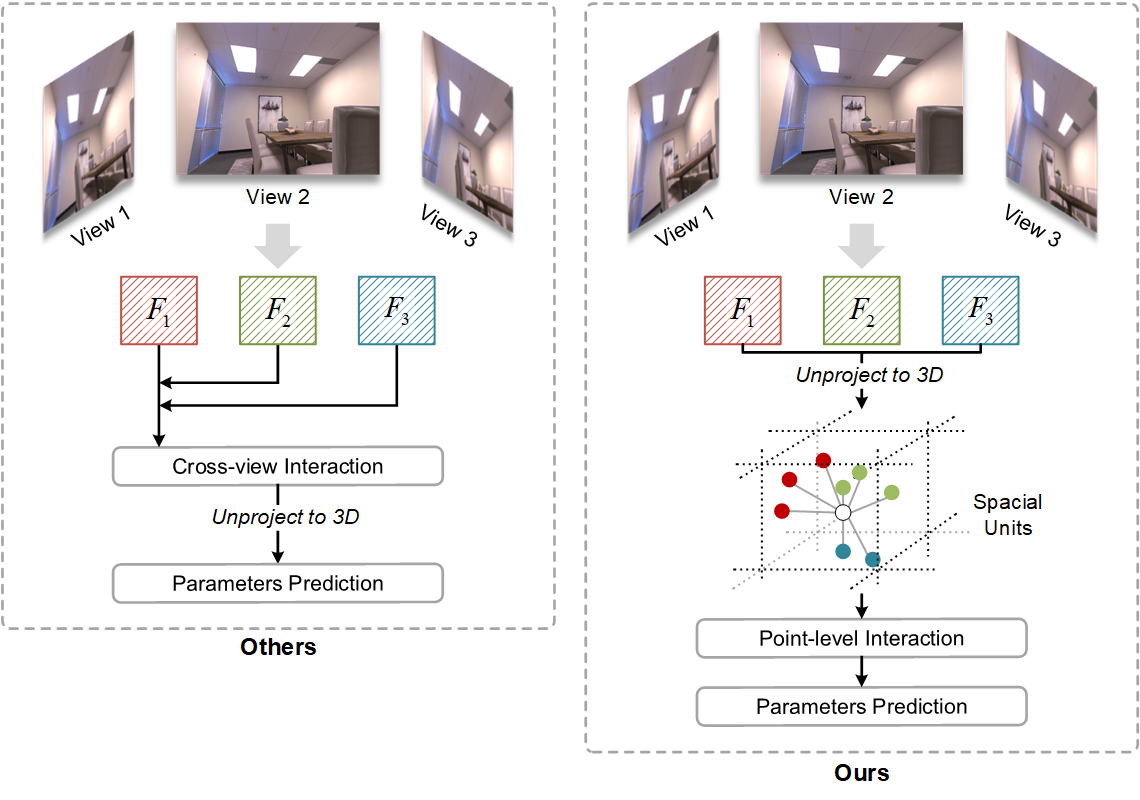}}
	\caption{\textbf{Point-level interaction module.} The left is the cross-view interaction method based on 2D features in other multi-view reconstruction approaches, and the right presents the process of our point-level interaction through space unit aggregation. }
   \vspace{-0.4cm}
    \label{fig_interaction}
\end{figure}

\subsubsection{Point-level Interaction}
Unlike other methods that perform cross-view interaction in 2D features, we propose first mapping multi-view features to 3D points and then achieving point-level interaction within spatial units, as shown in Figure \ref{fig_interaction}. The Gaussian distribution centers $\mu_0 \in\mathbb R^N$ are initialized by unprojecting the depth maps per pixel to the 3D space with camera poses. For the data where depth maps are unavailable, we employ CasMVSNet \cite{gu2020cascade} with similar architecture to the previous work \cite{chou2024gsnerf} to predict the depth of source views. We map $F'_r$ and $F'_s$ to $\mu_0$ as the initial point features and carry out point-level information interaction in the current space to further learn Gaussian attributes. Take semantic learning as an example, $\mu_0$ is regarded as a point cloud with original features $V_i (i\in N)$ and divided into local units at a fixed interval. The fusion function $\Omega$ incorporates the aggregated neighboring features of $x_i$ in the unit space into $V_i$:
\begin{align}
V_i^{'}&=\Omega(V_i, AvgPool(\{V_l\})\times Dis(x_i,x_l))\label{eq1}
\end{align}
where $\{V_l\}$ denotes the collection of the unit features corresponding to the $i$-th point, $x_l$ indicates the center point, $AvgPool$ is the average pooling function, $Dis$ is the distance from $x_i$ to the unit center. This point-level interaction on local space is capable of conducting efficient computations in large-scale Gaussian point clouds. The enhanced features with cross-view perspectives are utilized to improve Gaussian attributes for more flexible reconstruction. 

\begin{figure}[t]
        \centering{\includegraphics[width=1.0\linewidth]{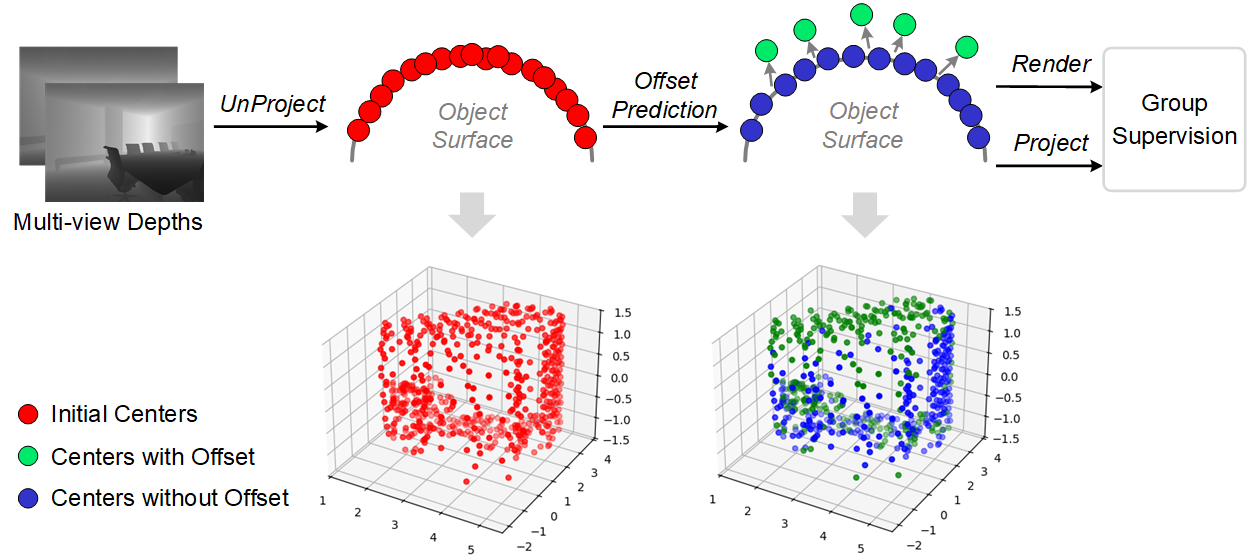}}
	\caption{\textbf{Offset learning module.} We divide the initial Gaussian centers (red) into two groups, one with offset (green) and one without offset (blue), and supervise them from both the rendering and geometric projection directions.}
   \vspace{-0.4cm}
    \label{fig_offset}
\end{figure}

\subsubsection{Gaussian Offset Learning}
The initial Gaussian points derived from multi-view depth maps are merely associated with the geometric surfaces of the visible portion. Rendering high-quality novel views requires learning more Gaussians to enrich the representation of 3D scenes. The pixel values in Gaussian splitting rendering are determined by the contribution of each Gaussian distribution in the observation direction, while for dense view reconstruction, the contributions of Gaussian points on geometric surfaces are redundant. Therefore, we propose a grouping supervision method based on the existing Gaussian centers for offset learning, as shown in Figure \ref{fig_offset}. The Gaussian centers obtained through back projection will cluster on the surface of objects. We predict offset probability $\hat{t}\in\mathbb R^N$ and distance $t\in\mathbb R^{N\times3}$ of original centers from $V_i^{'}$, which is aggregated with cross-view information. Consequently, the Gaussian distributions are partitioned into two groups, with one group still situated on the object surfaces, and the other moving into the local space outside the objects. After the offset prediction, the Gaussian means $\mu_i$ are updated as:
\begin{align}
\mu_i&=x_i+t_i\times Mask(\hat{t_i})\label{eq2}
\end{align}
where $Mask()$ selects points with a probability greater than the priori threshold (0.5). Center points that require offsets are used to create higher-quality images, while others without offsets still express the fundamental geometric properties of objects. We impose constraints on the offset prediction process to ensure that points without offsets can adequately describe the positions of object surfaces in space. The projective depth invariance of visible views is utilized for supervision:
\begin{align}
L_f&=\frac{1}{K}\sum_{v=1}^{K}smooth_{L1}(project(P_o,v),D_v)\label{eq3}
\end{align}
where $P_o$ represents the Gaussian centers without offset, $project(P_o,v)$ indicates the projected depth values of $P_o$ in the direction of view $v$, and $D_v$ denotes the original depth of the source views. 

\subsubsection{Rendering}
We predict the Gaussian parameters from the final point features ${V_i^{'}}$. Consider the semantic Gaussian model as an example, the predicted parameters can be expressed as follows:
\begin{align}
\mathcal P_s&=\{(\hat{t_i},t_i,S_i,R_i,s_i,\alpha_i)\}_{i \in N} \label{eq4}
\end{align}
where $s_i\in\mathbb R^\eta$ represents the semantic values with $\eta$ classes. Based on the aforementioned calculations, we acquire the ultimate expression of the Gaussian radiance fields $\mathcal G_s=(\mu,\Sigma,s,\alpha)$. Our generalizable semantic Gaussian Splatting directly models the scene as $\mathcal G_s$ without extra training and renders the semantic score maps $M_r \in \mathbb R^{H\times W \times \eta}$ for the novel views. Each score indicates the likelihood of a semantic class, and the highest scores constitute the results of semantic segmentation. 

\subsection{Training}
We construct the loss of the multi-task framework from the 3 aspects of color, semantics, and geometry constraint. We utilize Mean Squared Error (MSE) and Learned Perceptual Image Patch Similarity (LPIPS) to calculate the color construction loss $L_{C}$:
\begin{align}
L_C&=\lambda_1 MSE(I_r,I_g)+LPIPS(I_r,I_g) \label{eq5}
\end{align}
where $I_r$ is the rendered RGB image, $I_g$ is the target RGB image, $\lambda_1$ is the weight coefficient for MSE loss. The semantic loss $L_{S}$ is composed of two components about rendered semantics and multi-view semantics with cross-entropy loss function $L_{cross}$:
\begin{align}
L_S&=L_{cross}(M_r,\hat{M_r})+\sum_{v=1}^{K}L_{cross}(M_v,\hat{M_v}) \label{eq6}
\end{align}
Where $M_r$ and $\hat{M_r}$ represent the rendered semantics and classification labels of the unseen viewpoint, and $M_v$ and $\hat{M_v}$ denote the semantic segmentation results from 2D multi-view encoding and labels.

The total loss is defined as follows:
\begin{align}
Loss&= L_C+L_S+\lambda_f L_f+L_D \label{eq7}
\end{align}
Where $\lambda_f$ is the weight coefficient for geometric invariance loss $L_f$ in offset prediction. When the depth estimation network produces the depth maps, the depth loss $L_D$ is calculated using a smooth L1 loss function on the predicted depths and ground-truth values. Since the construction of Gaussians is highly dependent on the multi-view depths, the depth estimation network is learned first during the training stage, and then other components are incorporated for training. 

\begin{table*}[!t]
\begin{center}
	\caption{Comparison with recent generalizable methods on ScanNet Dataset}
    \renewcommand\arraystretch{1.3}
    \tabcolsep=4.5pt
    \label{tab1}
    \begin{threeparttable}
	\begin{tabular}{c|c|c|c|c|cccccc|c}
    \Xhline{2pt}
	  View&Methods &Year&Radiance Field &Depth & mIoU$\uparrow$ &acc$\uparrow$  &class acc$\uparrow$ &PSNR$\uparrow$ &SSIM$\uparrow$ &LPIPS$\downarrow$ &Time(s) \\
       \cline{1-12}  
        \multirow{7}{*}{K=8}&Neuray$^{\star}$\cite{liu2022neural} &2022 &NeRF &gt &52.09 &67.81 &61.98 &25.01 &83.07 &31.63 &-\\  
        &S-Ray \cite{liu2023semantic} &2023 &NeRF &gt  &55.53 &77.79 &60.92 &25.19 &83.66 &30.98 &7.19\\
        &\textbf{GSsplat}&- &3DGS &gt  &\textbf{64.35} &\textbf{87.91} &\textbf{70.81} &\textbf{28.00} &\textbf{90.33} &\textbf{13.97} &\textbf{0.22}\\
        \cline{2-12}  					
        &GeoNeRF$^{\star}$ \cite{johari2022geonerf} &2022&NeRF &mvs &53.78 &76.18 &61.90 &\textbf{32.55} &\textbf{90.88} &12.69 &-\\   		
        &GSNeRF \cite{chou2024gsnerf} &2024 &NeRF &mvs &58.30  &79.79 &65.93 &31.33 &90.73 &\textbf{12.53} &9.62\\	  					
        &MVsplat$^{\star}$ \cite{chen2025mvsplat} &2024 &3DGS &mvs  &47.56 &77.30 &56.56 &21.11 &81.62 &21.95 &0.61 \\	
        &\textbf{GSsplat} &- &3DGS &mvs &\textbf{60.38} &\textbf{82.01} &\textbf{66.96} &27.96 &90.24 &14.67 &\textbf{0.48}\\
        \Xhline{2pt}						
        \multirow{3}{*}{K=4}&GSNeRF &2024 &NeRF &mvs &53.42  &72.94 &62.73 &\textbf{32.23} &\textbf{93.75} &\textbf{11.29} &5.36\\	  		
        &MVsplat$^{\star}$ &2024 &3DGS &mvs  &48.19 &76.47 &56.84 &26.22 &87.88 &16.45 &0.26 \\													
        &\textbf{GSsplat} &- &3DGS &mvs &\textbf{59.51}\textcolor{red}{(6.09)} &\textbf{78.80}\textcolor{red}{(2.33)} &\textbf{66.97}\textcolor{red}{(4.24)} &30.25\textcolor{green}{(1.98)} &92.24\textcolor{green}{(1.51} &11.84\textcolor{green}{(0.55)} &\textbf{0.23}\\
        \cline{1-12} 												
        \multirow{3}{*}{K=2}&GSNeRF &2024 &NeRF &mvs &51.56  &73.03 &61.30 &\textbf{32.30} &\textbf{93.85} &10.69 &3.74 \\	  					
        &MVsplat$^{\star}$ &2024 &3DGS &mvs  &46.43 &75.34 &54.99 &28.14 &91.08 &12.70 &\textbf{0.10}\\										
        &\textbf{GSsplat} &- &3DGS &mvs &\textbf{57.57}\textcolor{red}{(5.92)} &\textbf{80.30}\textcolor{red}{(4.96)} &\textbf{65.54}\textcolor{red}{(4.24)} &31.12\textcolor{green}{(1.18)} &92.95\textcolor{green}{(0.9)} &\textbf{10.53}\textcolor{red}{(0.16)} &0.12 \\
    \Xhline{2pt}
	\end{tabular}  
     The highest item for each indicator is bolded. $ ^*$ represents that an extra semantic rendering module is incorporated into the original generalizable radiance field method for fair comparison. In the "Depth" column, "gt" represents using ground-truth depth, and "mvs" represents using MVSNet variants for depth estimation. The results with "K=8" of the radiance field methods based on NeRF are derived from \cite{chou2024gsnerf}. 
 \end{threeparttable}
\end{center}
\vspace{-0.4cm}
\end{table*}

\section{Experiment}
\label{sec:exp}
\subsection{Datasets and Metrics}
We train and evaluate our models using the same dataset settings as in previous work \cite{liu2023semantic, chou2024gsnerf}. The dataset of the real world is sourced from ScanNet \cite{dai2017scannet}, a large-scale indoor dataset that provides RGB-D scanned images, camera poses, and semantic annotations for 1513 scenes. For a fair comparison, we use 60 scenes for training and evaluate our generalizable Gaussian Splatting method on 10 unseen scenes. Replica \cite{straub2019replica} is a high-fidelity synthetic reconstruction dataset of indoor environments, composed of different video sequences from 18 scenes. For this task, 12 video sequences are utilized for training, and sampled frames from 4 new scene videos are used for evaluation. We integrate SSIM \cite{wang2004image}, PSNR, and LPIPS \cite{zhang2018unreasonable} as the image quality metrics. For semantic rendering, we utilize 3 common semantic segmentation metrics: pixel-wise classification accuracy (acc), mean accuracy for each category (class acc), and mean Intersection over Union (mIoU) for each category. 

\begin{table}[!t]
\begin{center}
	\caption{Comparison with recent methods on Replica Dataset}
    \renewcommand\arraystretch{1.3}
    \tabcolsep=5pt
    \label{tab2}
    \begin{threeparttable}
	\begin{tabular}{c|c|c|cc|c}
    \Xhline{2pt}
	  Methods &Radiance Field& Depth& mIoU$\uparrow$  &SSIM$\uparrow$ &Time(s) \\
       \cline{1-6}     
        Neuray$^{\star}$&NeRF &gt &44.37 &87.37 &- \\  
        S-Ray&NeRF &gt &45.30 &88.13 &6.69 \\
        \textbf{GSsplat}&3DGS&gt  &\textbf{55.14} &\textbf{91.42} &\textbf{0.22} \\
        \hline   
        GeoNeRF$^{\star}$&NeRF&mvs &45.12 &88.94 &-\\   		
        GSNeRF&NeRF  &mvs &51.52 &\textbf{92.44} & 9.55\\	 
        MVsplat$^{\star}$&3DGS &mvs &39.65 &74.03 &0.61\\	
        \textbf{GSsplat}&3DGS &mvs &\textbf{54.70}  &86.62 &\textbf{0.48}\\
    \Xhline{2pt}
	\end{tabular} 
    All settings are identical to those in Table \ref{tab1} with 8 views.
 \end{threeparttable}
\end{center}
\vspace{-0.4cm}

\end{table}
 \begin{figure*}[!b]
    \centering{\includegraphics[width=1.0\linewidth]{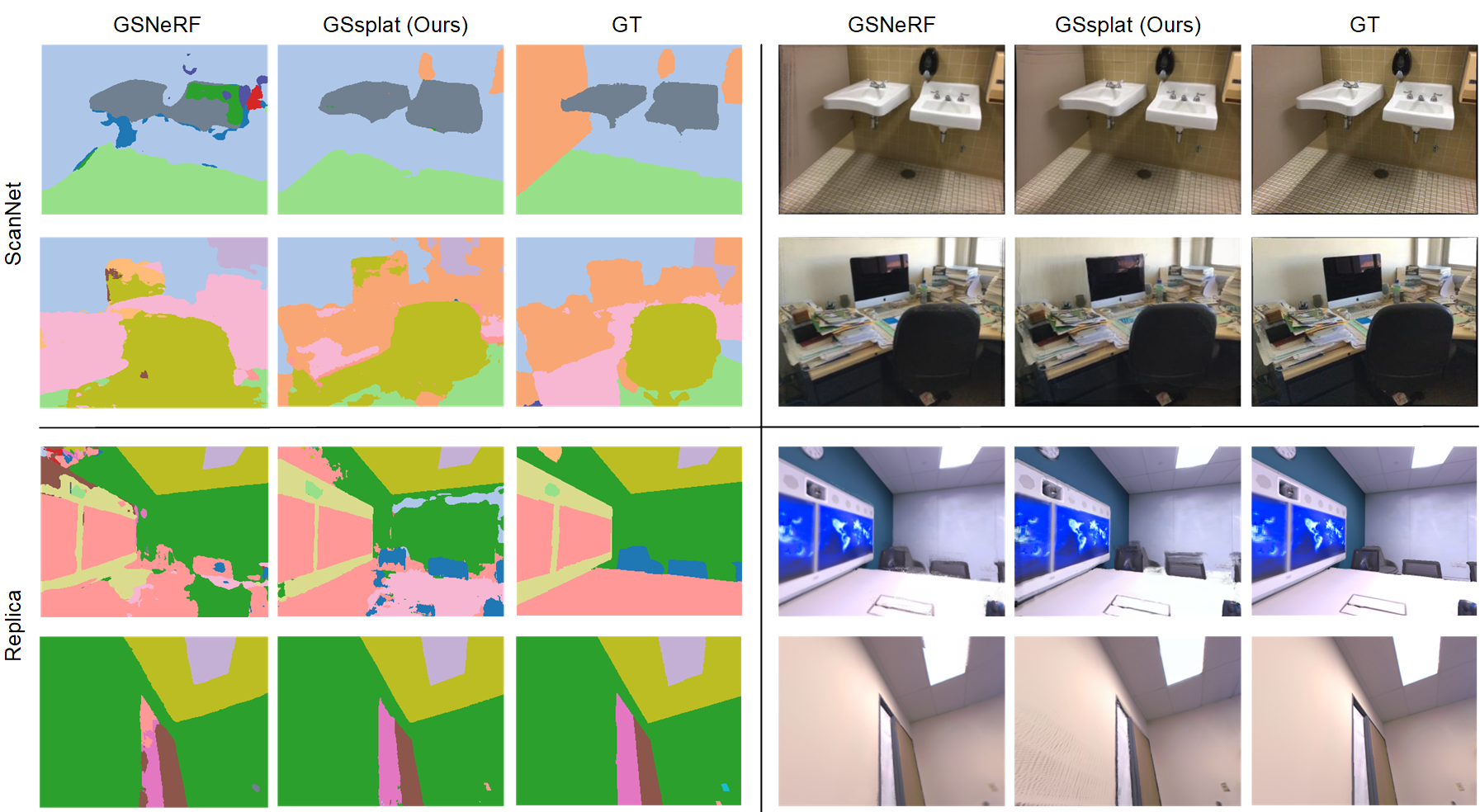}}
	\caption{\textbf{The demonstration of the novel-view synthesis.} Our method is compared with GSNeRF \cite{chou2024gsnerf} on 8-view inputs on evaluation datasets. The first two rows are from the test data of ScanNet, and the last two rows are from Replica.}
   \vspace{-0.4cm}
    \label{fig_res}
\end{figure*}

\subsection{Implementation Details}
The sizes of the source views and rendered images are both $320\times240$ pixels. The MSE loss weight coefficient $\lambda_1$ is set to $10.0$, and the offset loss weight coefficient $\lambda_f$ is set to $0.2$. All models are trained with a batch size of $2$ ($2\times K$ inputs for one batch) with the Adam \cite{kingma2014adam} optimizer. The leaning rate is set to $5 \times 10^{-4}$. When the multi-view depth is given, the training is conducted for 60,000 steps. When the depth estimation network is involved in the training, it is trained for 100,000 steps.  

\subsection{Comparison and Analysis}
We evaluate our method on the ScanNet and Replica datasets, comparing it with recent multi-task radiance field methods. The multi-view depth originates from ground truth or depth estimation networks of MVSNet variants, we conduct experiments for both of them. Since existing generalizable 3DGS-based methods do not support semantic synthesis on multi-view inputs, we integrate a semantic Gaussian branch that operates in parallel with the original Gaussian prediction into the most advanced method for fair comparison.

Our method is primarily compared to MVsplat \cite{chen2025mvsplat} and GSNeRF \cite{chou2024gsnerf}. MVsplat is a leading approach in generalizable 3DGS, demonstrating a high level of performance in color Gaussian reconstruction. GSNeRF, on the other hand, represents the latest research focusing on multi-task semantics and color synthesis, utilizing the generalizable NeRF framework. We analyze from the view synthesis effect and inference speed, especially the semantic understanding advantage of our method in generalizable radiance fields, and the reason why the 3DGS-based methods are inferior to NeRF-based methods in color synthesis.

\subsubsection{Semantic Rendering}
As shown in Table \ref{tab1}, we compare our GSsplat with the recent generalizable methods based on NeRF or 3DGS on the real-world dataset ScanNet. In the benchmark of 8 source views with unknown depths, we obtain the best results across all metrics for semantic synthesis, with a mIoU improvement of 2.08\% and an accuracy improvement of 2.22\% over the state-of-the-art NeRF-based method. Compared with the generalizable Gaussian method which uses the same depth estimation method, we have obtained obvious advantages in semantic segmentation results with 12.82\% in mIoU and 4.71\% in accuracy. It indicates that predicting semantic Gaussian with the same encoding structure as color Gaussian is inappropriate. Our method addresses this issue by processing the two features with different structures in deeper layers, and the 3D interaction occurs within their respective feature spaces before the Gaussian prediction. Table \ref{tab2} shows the main results of comparisons on the synthetic dataset Replica. Due to the limited training data consisting of only 8 scenes and unbalanced semantic categories, the overall semantic performance is lower than the results obtained from the 60-scene training dataset in ScanNet. Figure \ref{fig_res} shows our semantic rendering results compared with GSNeRF \cite{chou2024gsnerf}. It can be seen that our method is more precise in identifying semantic categories, despite similar segmentation contours.     

\subsubsection{Color Rendering}
In the evaluation from Table \ref{tab1} and Table \ref{tab2}, the color rendering quality indicators for the 3DGS-based methods are consistently lower than those for the NeRF-based methods. This is primarily due to the differences in rendering principles inherent in 3DGS and NeRF. The explicit reconstruction of 3DGS is based on geometric structure. The limitations of Gaussian distributions hinder their ability to manage complex lighting and scene noise. Consequently, even with entirely accurate input depth priors, they fail to achieve the same effect as dense sampling in NeRF. Semantic synthesis is less influenced by these factors, thus avoiding such issues. 

Our method shows significant superiority in terms of image quality indicators compared to the approaches based on 3DGS. This demonstrates that, within a generalizable framework utilizing deep priors, our method improves adaptability to complex scenes by learning Gaussian points that are not geometrically dependent. In addition, our method achieves comparable rendering effects to the NeRF-based method with only a 0.49\% disparity of SSIM score (90.24\%) from the most advanced NeRF-based method (90.73\%) on the ScanNet dataset. It is expected that better color rendering effects can be obtained by increasing the generalization training scale of the dataset.   

\begin{table}[!t]
\begin{center}
	\caption{The time required for each stage of the reasoning process.}
    \renewcommand\arraystretch{1.3}
    \tabcolsep=5pt
    \label{tab3}
    \begin{threeparttable}
	\begin{tabular}{c|c|c|cc|cc}
    \Xhline{2pt}
	  &Multi-view  &\multirow{2}{*}{Depth} &\multicolumn{2}{c|}{Semantic GS}  &\multicolumn{2}{c}{Color GS}  \\
      \cline{4-7} 
       &Network & &Create&Render&Create&Render\\
       \cline{1-7} 
       Time(s)&0.18 &0.26&0.02&0.003&0.02&0.006\\
    \Xhline{2pt}  
	\end{tabular} 
 \end{threeparttable}
\end{center}
\vspace{-0.6cm}
\end{table}

\begin{figure}[!b]
    \centering{\includegraphics[width=1.0\linewidth]{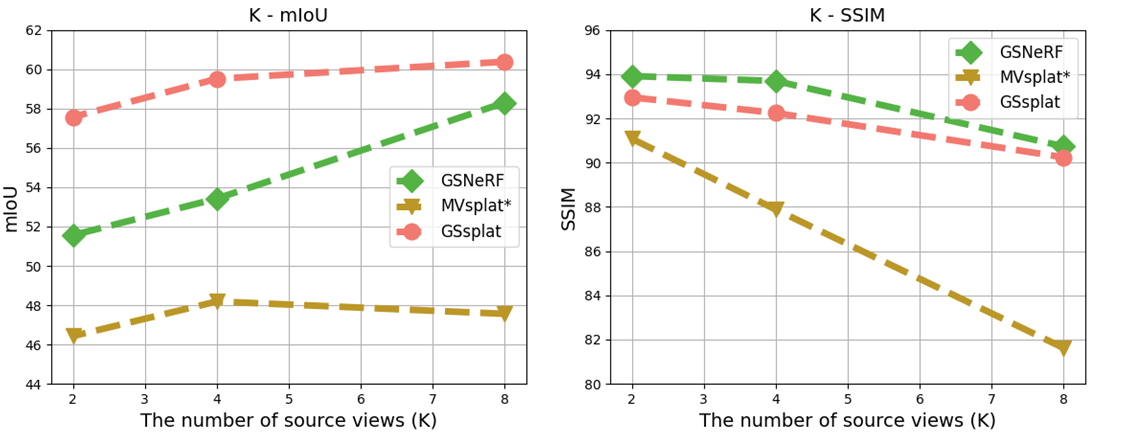}}
	\caption{Performance trend concerning reduced input views.}
   \vspace{-0.4cm}
    \label{fig_plot}
\end{figure}

\subsubsection{Feed-forward Speed}
In the last column of Table \ref{tab1} and Table \ref{tab2}, We present the time taken by each method to reconstruct the radiance fields from 8 multi-view inputs, along with multi-task rendering for a novel view. Our method attains the highest speed in all kinds of test scenarios. Specifically, it outperforms GSNeRF by more than 20 times with only 0.48 seconds in the test with depth estimation, and also achieves better real-time performance in comparison with 3DGS-based methods.

Does constructing Gaussian radiation fields for color and semantics separately reduce efficiency? The two Gaussian models share parameters in the multi-view encoding and Gaussian initialization stages. Due to differences in color and semantic features, we decoded and constructed two separate radiation fields. Table \ref{tab3} shows the time required for each stage of the reasoning process. Most of the time is consumed in multi-view network and depth estimation, which are computed once for one scene. Although two Gaussian Splatting models are created, they do not affect real-time performance. 

\subsubsection{Reduced Input Views}
To validate the robustness of our proposed generalizable Gaussian method, we conduct additional training and evaluation using reduced numbers of input views. Table \ref{tab1} shows the results and a comparison with the latest method. The semantic synthesis efficacy of our method with 4 or 2 source views still significantly outperforms others, and the image rendering quality is comparable to the NeRF-based method. Figure \ref{fig_plot} shows that our GSsplat consistently performs well, even with fewer views, while other methods exhibit considerable fluctuations. This is primarily because our method performs point-level interaction in 3D space, allowing different view information to direct the overall optimization of Gaussian parameters. 

Additionally, it is observed that all of these methods perform better in color rendering with fewer input views. The source views in this benchmark can effectively override the novel view, so adding more views yields little benefit for color mapping. The methods in the table rely on multi-view consistency as essential supervision during training. Fitting more views to the scene is a challenge for them.

\subsection{Ablation Studies}
We conduct extensive experiments to validate the effectiveness of the key components in our method. Firstly, we discuss the encoding structure in the hybrid network, which has strong pertinence for Gaussian reconstruction models with multiple tasks. Our network architecture is designed to achieve effective feature extraction for both color and semantics at the lowest cost. We also provide ablation experiments on the impact of the Gaussian offset learning and point-level interaction modules. Since our generalizable Gaussian Splatting does not concern the densification and pruning steps in scene-specific methods, these two modules can help our model adapt to the location and properties of Gaussians in different scenes.   

\begin{figure}[!b]
    \centering{\includegraphics[width=1.0\linewidth]{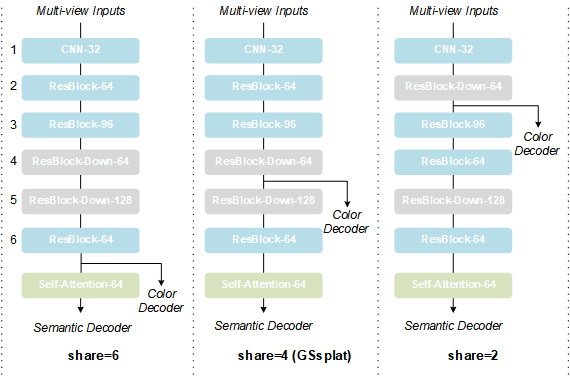}}
    \caption{The structure of our proposed multi-view network with different sharing layers. The gray blocks represent convolutional layers with stride=2. "share" indicates the number of blocks shared by the color encoder and the semantic encoder. The last digit in each block represents the number of convolutional kernels.}
   \vspace{-0.4cm}
    \label{fig_network}
\end{figure}

\begin{table*}[t]
\begin{center}
\caption{Ablation studies on the encoder structure of multi-view network}
    \renewcommand\arraystretch{1.5}
    \label{tab4}
    \begin{threeparttable}
	\begin{tabular}{cc|cccccc|cccccc}
      \Xhline{2pt}
      & &\multicolumn{6}{c|}{\textbf{ScanNet}} &\multicolumn{6}{c}{\textbf{Replica}}\\
      \cline{1-14} 
	  Share&Stride& mIoU$\uparrow$ &acc$\uparrow$  &class acc$\uparrow$ &PSNR$\uparrow$ &SSIM$\uparrow$ &LPIPS$\downarrow$ & mIoU$\uparrow$ &acc$\uparrow$  &class acc$\uparrow$ &PSNR$\uparrow$ &SSIM$\uparrow$ &LPIPS$\downarrow$\\
       \cline{1-14}  														
        2&2  &58.19 &84.80 &65.44 &27.57 &90.22 &14.07&51.30 &81.07 &59.65 &27.78 &91.09 &10.94 \\	
        4&2 &\textbf{64.35} &\textbf{87.91} &\textbf{70.81} &\textbf{28.00} &\textbf{90.33} &\textbf{13.97} &\textbf{55.14} &\textbf{82.67} &\textbf{63.00} &\textbf{28.47} &\textbf{91.42} &\textbf{10.47}\\
        6&4 &62.53 &84.92 &70.17 &27.94 &90.19 &14.27&53.33 &81.69 &62.40 &27.57 &90.20 &11.36 \\
    \Xhline{2pt}  										
   \end{tabular}  					
   "Share" indicates the number of convolutional blocks shared by the color encoder and the semantic encoder. "Stride" denotes the overall stride of color encoding during the convolution process.  The highest item for each indicator is bolded.
 \end{threeparttable}
\end{center}
\vspace{-0.4cm}
\end{table*}

\begin{table*}[t]
\begin{center}
\caption{Ablation studies on the self-attention layers.}
    \renewcommand\arraystretch{1.5}
    \label{tab5}
    \begin{threeparttable}
	\begin{tabular}{c|cccccc|cccccc}
      \Xhline{2pt}
       &\multicolumn{6}{c|}{\textbf{ScanNet}} &\multicolumn{6}{c}{\textbf{Replica}}\\
      \cline{1-13} 
	  Att-num& mIoU$\uparrow$ &acc$\uparrow$  &class acc$\uparrow$ &PSNR$\uparrow$ &SSIM$\uparrow$ &LPIPS$\downarrow$ & mIoU$\uparrow$ &acc$\uparrow$  &class acc$\uparrow$ &PSNR$\uparrow$ &SSIM$\uparrow$ &LPIPS$\downarrow$\\  					
       \cline{1-13}  																	
        1  &58.65 &82.98 &65.53 &27.67 &90.15 &14.06 &49.84 &81.03 &58.44 &27.87 &91.09 &10.49 \\	
        2  &62.05 &83.19 &68.41 &27.19 &89.79 &14.63&52.32 &81.78 &62.57 &27.93 &91.36 &10.67 \\
        3 &\textbf{64.35} &\textbf{87.91} &\textbf{70.81} &\textbf{28.00} &\textbf{90.33} &\textbf{13.97} &\textbf{55.14} &\textbf{82.67} &63.00 &28.47 &91.42 &10.47\\  					
        4&62.19	 &85.30 &69.41 &27.34 &90.08 &14.15&55.02 &82.32 &\textbf{63.67} &\textbf{28.50} &\textbf{91.79} &\textbf{9.29} \\ 		
    \Xhline{2pt}  															
   \end{tabular}  					
   "Att-num" indicates the number of self-attention layers.  The highest item for each indicator is bolded.
 \end{threeparttable}
\end{center}
\vspace{-0.6cm}
\end{table*}
\begin{figure*}[!b]
    \centering{\includegraphics[width=1.0\linewidth]{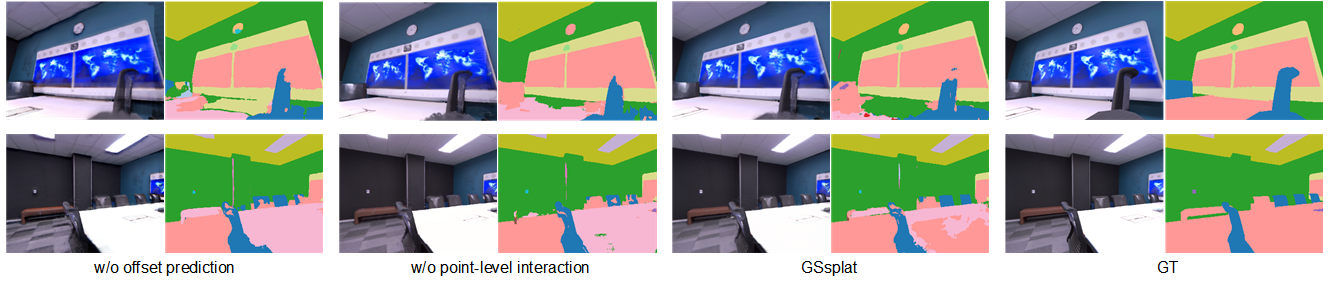}}
	\caption{Results of 2 novel-views of one of the Replica test scenes conducted during the ablation studies. The first column in each block is the rendered color images, and the second column is the semantic results.}
    \label{fig_ablation}
\end{figure*}
\subsubsection{Multi-view Network}
The most important is the position of the color decoder node within the branch, which determines whether a multi-task network can effectively extract information from different feature spaces. We test 3 structures as shown in Figure \ref{fig_network}, corresponding to the number of sharing blocks of 6, 4, and 2. Our GSsplat uses the middle structure, where color and semantic encoding share the first 4 blocks, each containing 2 convolutional layers. Table \ref{tab4} shows the synthesis results on ScanNet and Replica datasets. It can be observed that an overly shallow or deep shared network will lead to a decline in segmentation performance. In terms of color rendering, the deepening of the encoding network did not bring about obvious enhancement to the image quality results, indicating that colors are more intuitive and easier to learn than semantics. 

Table \ref{tab5} presents the evaluation results from two datasets, featuring a varying number of self-attention layers in the semantic encoder. With the increase in the number of self-attention layers, the novel-view synthesis effect from the generalization method, especially in semantic synthesis, is greatly improved. However, there is no obvious benefit after the number of layers increases to 4. Based on both performance and time considerations, our GSsplat sets the number of self-attention layers to 3. 

\begin{table}[!t]
\begin{center}
	\caption{Ablation studies of key components }
    \renewcommand\arraystretch{1.3}
    \label{tab6}
    \begin{threeparttable}
	\begin{tabular}{c|cccc}
    \Xhline{2pt}
	   &mIoU$\uparrow$  &acc$\uparrow$ & PSNR$\uparrow$  &SSIM$\uparrow$ \\
       \cline{1-5} 					
        \textbf{GSsplat}&\textbf{55.14} &\textbf{82.67} &\textbf{28.47} &\textbf{91.42}  \\ 		
        w/o offset prediction &51.39 &80.05 &25.57 &88.53 \\
        w/o offset supervision &52.58&82.24 &27.79 &91.36 \\
        w/o interaction &53.03 &81.07 &28.16 &91.17 \\
    \Xhline{2pt}  
	\end{tabular} 
 \end{threeparttable}
\end{center}
\vspace{-0.6cm}
\end{table}

\subsubsection{Gaussian Offset Learning}
The ablations in Table \ref{tab6} are based on the Replica dataset with ground truth depths. Replica is a synthetic dataset, and its depth maps are less noisy and more complete than those from ScanNet, which avoids the uncertainty associated with multi-view depth in the inference process. In Table \ref{tab6}, "w/o offset prediction" indicates removing the entire Gaussian offset learning module from GSsplat, and "w/o offset supervision" implies predicting Gaussian offset but removing the extra supervision. From the results of the first three rows, it can be observed that the Gaussian offset learning module exerts a considerable influence on constructing both semantic Gaussians and color Gaussians. This manifests the difference between geometric points from depth maps and Gaussian points for high-quality rendering. Our group-supervision method can adaptively acquire this difference without supplementary Gaussian distributions. 

Table \ref{tab7} shows the proportion of Gaussian centers with offset in reconstruction for each scene in evaluation datasets. The last two columns represent the mean perception of multiple 8-view inputs on color and semantic Gaussians. To better illustrate the relationship between the offset points and the geometry of the scene, we generate the initial Gaussian positions using precise depth priors. It can be observed that most Gaussian radiance fields are reconstructed with more than half of the Gaussian distributions being offset, which further highlights the significance of our offset learning module. 

\begin{table}[!t]
\begin{center}
	\caption{The proportion of the Gaussian centers with offset in color and semantic Gaussians}
    \renewcommand\arraystretch{1.3}
    \tabcolsep=8pt
    \label{tab7}
    \begin{threeparttable}
	\begin{tabular}{c|c|cc}
    \Xhline{2pt}
	  Source&Scene Name&Color GS &Semantic GS \\
       \cline{1-4} 									
         \multirow{2}{*}{Replica} &office\_4&80.99\%	&55.18\%\\
        &room\_2	&88.91\%	&68.10\%\\
        \cline{1-4} 
        \multirow{10}{*}{ScanNet}&scene0063\_00	&70.64\%	&58.76\%\\
        &scene0067\_00	&67.52\%	&50.81\%\\
        &scene0071\_00	&84.87\%	&53.59\%\\
        &scene0074\_00	&55.16\%	&50.12\%\\
        &scene0079\_00	&46.81\%	&52.75\%\\
        &scene0086\_00	&53.34\%	&52.92\%\\
        &scene0200\_00	&49.84\%	&52.23\%\\
        &scene0211\_00	&44.00\%	&47.53\%\\
        &scene0226\_00	&49.28\%	&60.26\%\\
        &scene0376\_02	&66.28\%	&60.55\%\\
    \Xhline{2pt}
	\end{tabular}
 \end{threeparttable}
\end{center}
\vspace{-0.4cm}
\end{table}

\begin{table}[!t]
\begin{center}
	\caption{The Impact of Joint Training on Gaussian Reconstruction}
    \renewcommand\arraystretch{1.3}
    \label{tab8}
    \begin{threeparttable}
	\begin{tabular}{c|cccc}
    \Xhline{2pt}
	  3D Gaussian &mIoU$\uparrow$  &acc$\uparrow$ & PSNR$\uparrow$  &SSIM$\uparrow$ \\
       \cline{1-5} 					
        semantic&56.03 &80.79 &- &- \\ 		
         color &- &- &28.02 &90.73 \\
        semantic+color &60.38&82.01 &27.96 &90.24 \\
    \Xhline{2pt}  
	\end{tabular} 
 \end{threeparttable}
\end{center}
\vspace{-0.6cm}
\end{table}

\subsubsection{Point-level Interaction}
The last row labeled "w/o interaction" in Table \ref{tab6} presents the results without the point-level interaction module. It demonstrates that the local feature aggregation in spatial units exerts a considerable influence on the reconstruction and rendering of the semantic Gaussian radiance field. Most of the current methods incorporate 2D semantic segmentation supervision into training, but the semantic information extracted from a single view is not sufficient for 3D semantic understanding. Our method allows Gaussians to learn semantic-related geometric information from neighboring points, enhancing scene semantic understanding. Figure \ref{fig_ablation} presents the rendering results of an unseen scene devoid of the important modules.

\subsubsection{Joint Training}
In our multi-task framework, the semantic and color Gaussian reconstruction models are trained jointly. They share view encoding and initial geometry, decoding Gaussian parameters in different feature spaces. Will these influence each other? We verify it in Table \ref{tab8}, the top 2 rows present synthesis results on ScanNet from separately trained color and semantic models. For color rendering, the image quality from the joint training model is comparable to that of the independent training model. The semantic Gaussian radiation field of independent training is slightly lower than that of joint training, suggesting that color Gaussian learning significantly aids semantic understanding. The joint training framework enhances view synthesis quality and lowers computation costs.

\section{Conclusion}
In this paper, we introduce a generalizable semantic Gaussian Splatting method (GSsplat) for 3D scene understanding. It can efficiently reconstruct color and semantic Gaussians from multi-view inputs and synthesize novel-view images and semantic maps. We design a universal hybrid network as the backbone of our multi-task framework, which extracts semantic and color information separately at various levels. We focus on improving rendering effects by non-geometric alignment of Gaussian positions and propose a novel offset learning module with group-based supervision. Unlike existing cross-view interaction methods that rely on 2D visual encoding, multi-view features are aggregated in 3D space using a point-level interaction module for geometry-related learning. Our GSsplat is evaluated on ScanNet and Replica datasets and outperforms the state-of-the-art methods on semantic segmentation results. We also achieve the fastest free-forward speed compared to other multi-task generalizable methods, especially much faster than NERF-based ones. Although the explicit nature of 3DGS leads to inferior color rendering to NeRF, our image quality exceeds other 3DGS-based methods.

\section*{Acknowledgments}
This work presented in this paper is partially supported by grant from Fundamental Research Funds for the Central Universities(No. 2024ZYGXZR104).

\bibliographystyle{IEEEtran}
\bibliography{literature}

\end{document}